\documentclass[traditabstract]{aa}
\usepackage{graphicx}
\usepackage{txfonts}
\newcommand{\cflame}{Combustion and Flame}         
\newcommand{\ijck}{Int. J. Chem. Kinet.}           
\newcommand{\nature}{Nature}                       
\newcommand{\pss}{Planet. Space Sci.}              

\begin{document}
\title{The impact of atmospheric circulation on the chemistry of the hot Jupiter HD 209458b}
\titlerunning{Impact of atmospheric circulation on the chemistry of HD 209458b}
\authorrunning{Ag\'undez et al.}

\author{M. Ag\'undez\inst{1,2}, O. Venot\inst{1,2}, N. Iro\inst{3}, F. Selsis\inst{1,2}, F. Hersant\inst{1,2}, E. H\'ebrard\inst{1,2}, and M. Dobrijevic\inst{1,2}}

\institute{Univ. Bordeaux, LAB, UMR 5804,
F-33270, Floirac, France \and CNRS, LAB, UMR 5804, F-33270, Floirac, France \and Astrophysics Group, Keele University, Keele, Staffordshire, ST5 5BG, UK\\
\email{Marcelino.Agundez@obs.u-bordeaux1.fr}}

\date{Received; accepted}


\abstract
{We investigate the effects of atmospheric circulation on the
chemistry of the hot Jupiter HD 209458b. We use a simplified
dynamical model and a robust chemical network, as opposed to
previous studies which have used a three dimensional circulation
model coupled to a simple chemical kinetics scheme. The
temperature structure and distribution of the main atmospheric
constituents are calculated in the limit of an atmosphere that
rotates as a solid body with an equatorial rotation rate of 1 km
s$^{-1}$. Such motion mimics a uniform zonal wind which resembles
the equatorial superrotation structure found by three dimensional
circulation models. The uneven heating of this tidally locked
planet causes, even in the presence of such a strong zonal wind,
large temperature contrasts between the dayside and nightside, of
up to 800 K. This would result in important longitudinal
variations of some molecular abundances if the atmosphere were at
chemical equilibrium. The zonal wind, however, acts as a powerful
disequilibrium process. We identify the existence of a pressure
level of transition between two regimes, which may be located
between 100 and 0.1 mbar depending on the molecule. Below this
transition layer, chemical equilibrium holds, while above it, the
zonal wind tends to homogenize the chemical composition of the
atmosphere, bringing molecular abundances in the limb and
nightside regions close to chemical equilibrium values
characteristic of the dayside, i.e. producing an horizontal
quenching effect in the abundances. Reasoning based on timescales
arguments indicates that horizontal and vertical mixing are likely
to compete in HD 209458b's atmosphere, producing a complex
distribution where molecular abundances are quenched horizontally
to dayside values and vertically to chemical equilibrium values
characteristic of deep layers. Either assuming pure horizontal
mixing or pure vertical mixing, we find substantial variations in
the molecular abundances at the evening and morning limbs, up to
one order of magnitude for CH$_4$, which may have consequences for
the interpretation of transmission spectra that sample the
planet's terminator of hot Jupiters.}
{}
{}
{}
{}


\maketitle

\section{Introduction}

Gas giant planets that orbit close to their star, the so-called
hot Jupiters, are not the most common type of extrasolar planets
(\cite{bat2012} 2012), although they are the easiest to observe
due to their short orbital distance and large mass. The first
planets discovered around main sequence stars other than the Sun
are in fact hot Jupiters, such as 51 Pegasi b, the first such
object found (\cite{may1995} 1995), and HD209458b, the first
exoplanet caught transiting its star (\cite{cha2000} 2000;
\cite{hen2000} 2000). Moreover, hot Jupiters are nearly the only
ones among extrasolar planets for which constraints on their
atmospheric composition have been obtained from observations.
Transiting exoplanets offer indeed the opportunity to get
transmission and emission spectra of the atmosphere by observing
at different wavelengths during the primary transit and secondary
eclipse, respectively. The interpretation of such spectra,
although difficult and sometimes contradictory, allows to identify
and get the abundances of the main atmospheric constituents
(\cite{cha2002} 2002; \cite{tin2007} 2007; \cite{swa2008} 2008,
2009; \cite{gri2008} 2008; \cite{sin2009} 2009; \cite{mad2011}
2011; \cite{bea2011} 2011). Incoming missions such as FINESSE,
James Webb Space Telescope, and EChO will in the near future allow
to observe the atmospheres of transiting exoplanets down to the
size of rocky planets, although to date hot Jupiters offer the
best and almost unique chance to get access to the atmospheric
composition of extrasolar planets.

The characterization of hot Jupiter atmospheres has motivated the
development of one dimensional models that include thermochemical
kinetics, diffusion, and photochemistry, and that aim at
describing the chemical behaviour and composition of such
atmospheres in the vertical direction (\cite{zah2009} 2009;
\cite{lin2010} 2010; \cite{mos2011} 2011; Kopparapu et al. 2012;
Venot et al. 2012). These models indicate that chemical
equilibrium is attained deep in the atmosphere, although the
chemical composition of the layers typically sampled by transit
observations, in the 1 bar to 0.01 mbar pressure regime, is
maintained out of equilibrium due to two major disequilibrium
processes. On the one hand, the intense ultraviolet radiation
received from the star on top of the atmosphere drives an active
photochemistry which may extend down to the 1 mbar layer. On the
other, vertical mixing processes tend to homogenize the chemical
composition above a certain height as a consequence of the rapid
transport of material from deep and hot layers to higher and
cooler altitudes, where chemical kinetics is much slower.

Hot Jupiters are, according to theory (see e.g. Guillot et al.
1996), tidally locked to their star and thus receive stellar light
on one hemisphere only. Therefore, in the absence of atmospheric
winds, there would be a extremely high temperature contrast
between the dayside and the nightside. General circulation models,
however, predict the presence of strong winds, with velocities up
to a few km s$^{-1}$, that would efficiently redistribute the
energy from the dayside to the nightside (Guillot \& Showman 2002;
Cho et al. 2003, 2008; Cooper \& Showman 2005; Showman et al.
2008, 2009; Menou \& Rauscher 2009; Heng et al. 2011; Miller-Ricci
Kempton \& Rauscher 2012; Dobbs-Dixon et al. 2012).
Moreover, observational evidence of such winds has been
tentatively found by Snellen et al. (2010). These authors have
observed a 2 km s$^{-1}$ blueshift of carbon monoxide absorption
lines in the transmission spectrum of HD 209458b at the 2$\sigma$
confidence level, and have interpreted it as an evidence of
day-to-night winds occurring in the 0.01--0.1 mbar pressure
regime.

Winds transport material between different locations in the
atmosphere and, therefore, may have an impact on the distribution
of the atmospheric constituents. In a study that investigated this
phenomenon, Cooper \& Showman (2006) coupled a simplified kinetics
scheme for the conversion between CO and CH$_4$ to a three
dimensional circulation model of HD 209458b, and showed that
dynamics driven by uneven heating of the planet acts as a strong
disequilibrium process and homogenize the mixing ratios of CO and
CH$_4$ in the 1 bar to 1 mbar pressure range, even in the presence
of strong temperature gradients. It is, however, difficult to
disentangle from such complex circulation models whether is
vertical or horizontal transport the dominant disequilibrium
process. In their study, Cooper \& Showman (2006) argue in terms
of timescale estimates, and conclude that horizontal transport is
not important compared to vertical mixing.

Here we adopt a different approach to study the impact of
horizontal winds on the distribution of the atmospheric
constituents of HD 209458b. We use a time-dependent radiative
model to calculate the temperature structure of an atmosphere that
rotates as a solid body, mimicking a uniform zonal\footnote{The
terms zonal and meridional refer to the west-east and north-south
directions, respectively. In the case of tidally locked planets,
the substellar point is usually assigned longitude 0$^\circ$ and
eastward is positive (e.g. 0$^\circ$ $\rightarrow$ +90$^\circ$).
As with the Earth, the rotation of the planet occurs eastward.}
wind. Afterwards, the chemical evolution is computed at various
pressure levels with a robust chemical kinetics network to get the
chemical composition as a function of longitude and height.

\section{The model}

\subsection{The radiative model}

\begin{figure}
\centering \vspace{-0.40cm}
{\includegraphics[angle=0,width=1.085\columnwidth]{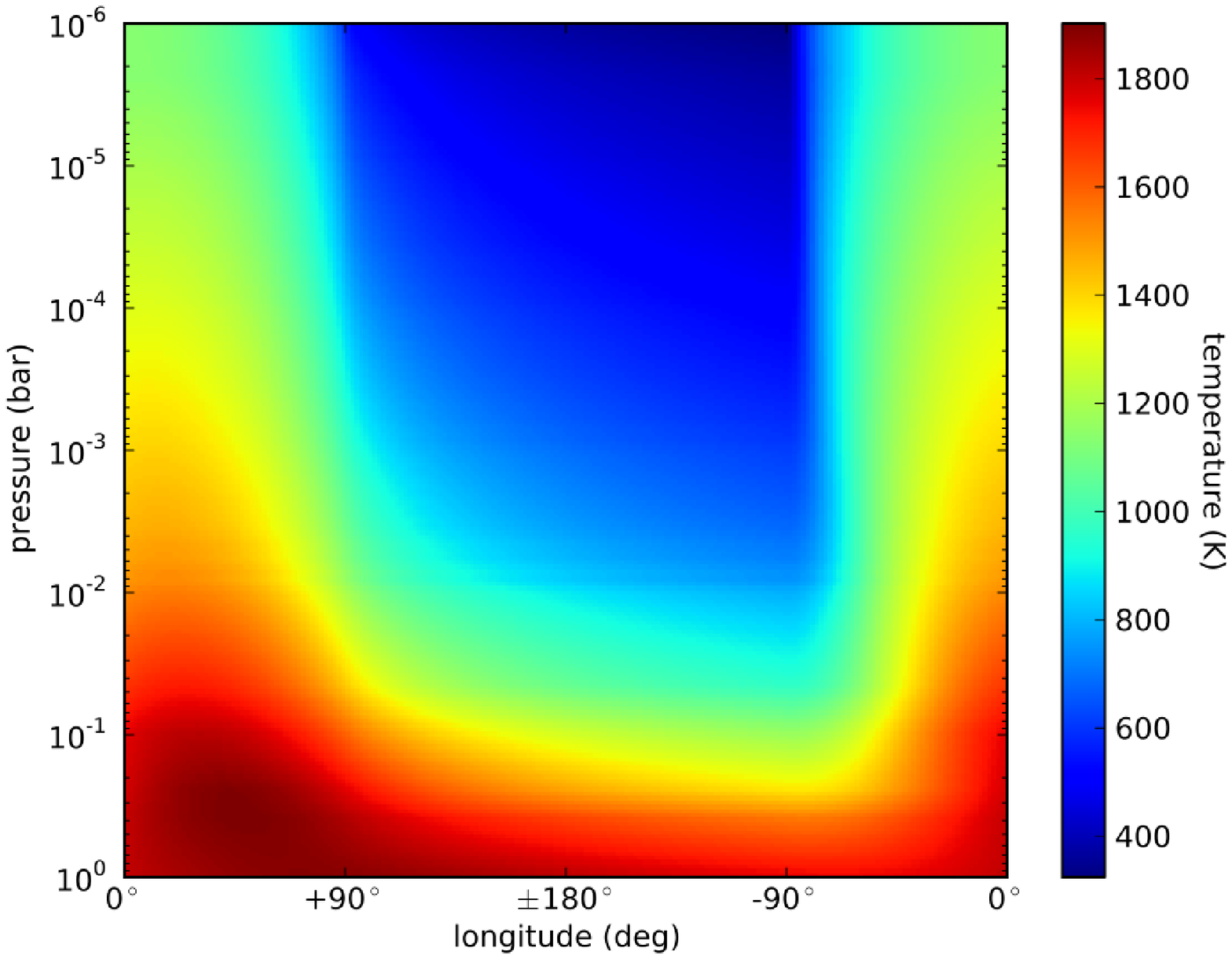}\vspace{0.35cm}}
{\includegraphics[angle=0,width=0.95\columnwidth]{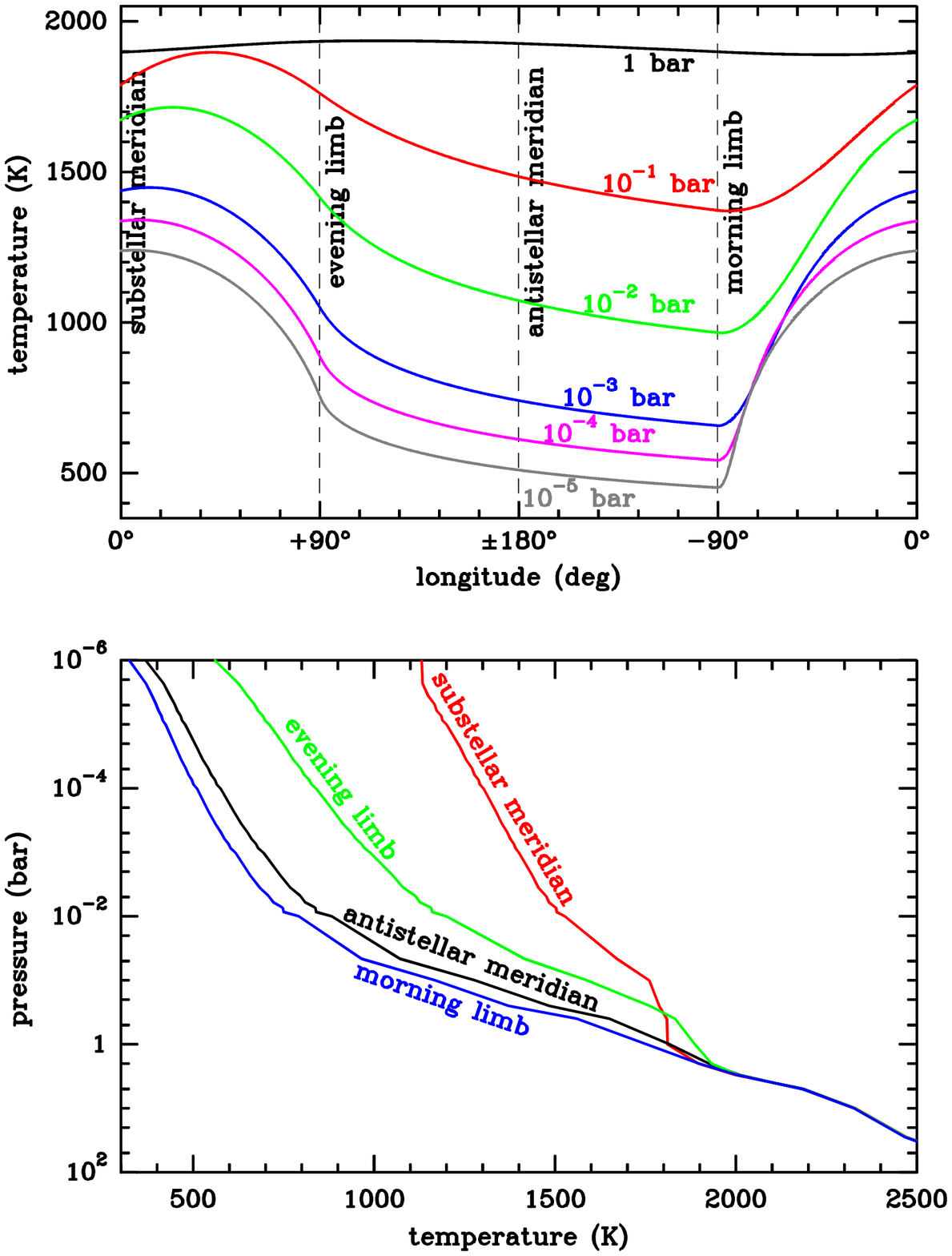}}
\caption{Thermal atmospheric structure of HD 209458b. The top
panel shows the temperature distribution as a function of
longitude and pressure, while the middle panel shows the
temperature as a function of longitude at selected pressure
levels. The bottom panel shows the vertical temperature profiles
at the substellar and antistellar meridians, and at the morning
and evening limbs.} \label{fig-tk-distribution}
\end{figure}

A detailed description of the adopted physical model is given in
Iro et al. (2005). Briefly, we consider a planet tidally locked to
its star, i.e. in synchronous rotation, with a radius of 1.38
R$_J$ and a mass of 0.714 M$_J$, as derived for the hot Jupiter HD
209458b (Southworth 2010). We consider that the atmosphere rotates
as a solid body, with respect to the synchronously rotating frame,
with a constant angular velocity corresponding to an equatorial
linear velocity of 1 km s$^{-1}$ at the 1 bar pressure level. Such
motion of the atmosphere mimics a uniform zonal wind independent
of height, where layers remain static with respect to each other.
Under this assumption, the evolution of the vertical temperature
profile is given by
\begin{equation}
\frac{dT}{dt} = \frac{\langle mg \rangle}{C_p} \Big( \frac{dF}{dp}
\Big) \label{eq-temperature-evolution}
\end{equation}
where $T$ is the temperature, $t$ the time, $\langle mg \rangle$
the mean molecular weight, $C_p$ the specific heat, $F$ the
radiative net flux, and $p$ the pressure. The incoming stellar
flux is computed adopting a Kurucz spectrum for HD
209458\footnote{See
\texttt{http://kurucz.harvard.edu/stars/hd209458/}} ($T_{\rm eff}$
= 6100 K and log $g$ = 4.38), a stellar radius of 1.2 R$_\odot$
(Mazeh et al. 2000), and an orbital distance of 0.047 AU
(Southworth 2010). The longitude-dependent insolation
pattern is calculated for a latitude of 30$^\circ$, which
corresponds approximately to the flux averaged over latitude, i.e.
along a meridian (see details in Iro et al. (2005). The sources
of atmospheric opacity included are Rayleigh scattering,
absorption from H$_2$--H$_2$ and H$_2$--He pairs, H$^-$
bound-free, H$_2^-$ free-free absorption, ro-vibrational lines of
the molecules CO, H$_2$O, CH$_4$, CO$_2$, NH$_3$, and TiO, and
resonance lines of the alkali atoms Na and K. The sources of the
opacity data used are described in Iro et al. (2005), except for
NH$_3$ and CO$_2$, whose spectroscopic data were taken from HITRAN
2008 and HITEMP respectively (Rothman et al. 2009, 2010). In the
current model, 45 layers are included with pressures ranging from
100 to 10$^{-6}$ bar. We adopt as initial condition a vertical
temperature profile computed at radiative equilibrium with a one
dimensional radiative-convective model under planet-averaged
insolation conditions. The abundances of the species that provide
opacity are calculated at chemical equilibrium, adopting solar
elemental abundances and the radiative equilibrium temperature
profile, and are assumed to remain constant with time, and
therefore with longitude (see Iro et al. 2005 for more details).
The thermal structure and chemical composition could be
calculated self-consistently by iterating between the radiative
and chemical models. In the case of HD 209458b our approach is,
however, justified by the fact that the main species that provide
opacity and affect the thermal structure are H$_2$O and CO, whose
abundances, as will be shown in Sec.~\ref{sec-results}, are very
close to the chemical equilibrium values and remain uniform with
longitude. The thermal structure of the atmosphere (temperature
as a function of longitude and height) is evaluated by integrating
Eq.~(\ref{eq-temperature-evolution}) during several rotation
cycles until the temperature reaches a periodic state in each
layer. We note that below the 10 bar level the temperature does
not completely reach a periodic state due to the very long
radiative timescale. These deep layers have, however, little
interest to study the variation with longitude of the chemical
composition since the temperature is nearly uniform with
longitude.

The resulting temperature distribution is shown in three different
ways in Fig.~\ref{fig-tk-distribution}. The increase in the
radiative timescale, or thermal inertia, with increasing depth
produces a couple of interesting effects. First, the maximum of
temperature is shifted in longitude with respect to the substellar
meridian by an amount that increases with depth (see top and
middle panels of Fig.~\ref{fig-tk-distribution} in the
1--10$^{-2}$ bar pressure regime), which is a simple
consequence of the eastward transport of heat. An eastward jet and
the resulting shift of the thermal emission peak was predicted by
Showman \& Guillot (2002) to be a common phenomenon in hot Jupiter
atmospheres. The shift was later observed in the thermal phase
curve of HD 189733b by Knutson et al. (2007). In the same vein,
we find that the temperature is not homogeneous in the nightside
(which would be the case in the absence of winds), but rather
inhomogeneous, with the coldest regions located close to the
morning limb (see mid and bottom panels of
Fig.~\ref{fig-tk-distribution}). A second effect worth to note is
that the longitudinal profile of temperature is markedly different
in the deep and high regions of the atmosphere. Below the 1 bar
pressure level the temperature remains nearly uniform (see bottom
panel of Fig.~\ref{fig-tk-distribution}), while above this level
large temperature contrasts, up to 800 K, exist (see middle panel
of Fig.~\ref{fig-tk-distribution}).

Our time-dependent radiative model provides an accurate treatment
of the radiative processes, although the atmospheric dynamics is
somewhat simplified. In the limit of a solid body rotating
atmosphere, the dynamics is restricted to a uniform zonal wind.
Meridional winds are, therefore, neglected as well as vertical
transport processes (layers are radiatively coupled but
dynamically decoupled). Focusing on horizontal advection (winds),
three dimensional circulation models of hot Jupiters (e.g. Cooper
\& Showman 2005, 2006; Showman et al. 2008, 2009; Heng et al.
2011) find that the atmospheric circulation is dominated by a fast
eastward (superrotating) jet stream that develops at low latitudes
and may reach wind speeds up to a few km s$^{-1}$. Equatorial
superrotation dominates the dynamics of the atmosphere in the deep
regions but as we move towards higher altitudes the circulation
pattern changes. Three dimensional circulation models of HD
209458b (see e.g. Showman et al. 2009, 2012) indicate that above
the 1--0.1 mbar level zonal winds are no longer superrotating and
strong substellar-to-antistellar zonal and meridional winds
dominate the atmospheric dynamics. Our assumption of a
uniform zonal wind with an equatorial velocity of 1 km s$^{-1}$
also ignores variations of the zonal wind speed with altitude,
which may be significant (wind speeds generally increase with
height) according to circulation models (e.g. Showman et al.
2009). With the above limitations, our approach should still
offer a realistic approximation of the atmospheric dynamics of HD
209458b in the 1--10$^{-4}$ bar pressure range, where equatorial
superrotation holds. This is in fact the most interesting region
to study the disequilibrium effects caused by atmospheric
circulation, as will be shown hereafter.

\subsection{The chemical model} \label{subsec-model-chemistry}

Still under the assumption of a solid body rotating atmosphere,
the evolution in the chemical composition of the different
atmospheric layers as they cycle around the planet is calculated
by solving a system of differential equations of the type
\begin{equation}
\frac{dy_i}{dt} = \frac{P_i - L_i}{n} \label{eq-abundance-evolution}
\end{equation}
where $y_i$ is the mole fraction of the species $i$, $P_i$ and
$L_i$ are the rates of production and loss of species $i$ due to
chemical reactions (with cgs units of cm$^{-3}$ s$^{-1}$), and $n$
is the total volume density of particles (with cgs units of
cm$^{-3}$), which is related to the pressure $p$ and temperature
$T$ through the ideal gas law. The system of chemical equations is
integrated, independently for each layer, during several rotation
cycles until the abundances of the major species reach a periodic
state. During the integration the pressure remains constant in
each layer and the temperature varies according to the
longitude-dependent profiles calculated with the radiative model.
We start the integration at the substellar meridian with a
chemical composition calculated at chemical equilibrium with solar
elemental abundances (Asplund et al. 2009) and the ($T, p$)
profile shown in the bottom panel of
Fig.~\ref{fig-tk-distribution}. The chemical equilibrium
calculations are done with a code that minimizes the Gibbs energy
following the algorithm of Gordon \& McBride (1994). The choice of
the initial composition at the chemical equilibrium values of hot
regions such as the substellar meridian results very convenient as
it accelerates considerably the convergence towards a periodic
state. The reasons for that will be discussed in
Sec.~\ref{subsec-chemistry-superrotation}.

The adopted chemical network, available from the KIDA
database\footnote{See
\texttt{http://kida.obs.u-bordeaux1.fr/models}} (Wakelam et al.
2012), is described in detail in Venot et al. (2012). Here we use
their nominal mechanism which consists of 102 neutral species
composed of C, H, O, and N (He is just a non reactive collider in
some reactions) linked by 1882 chemical reactions of three main
types: bimolecular disproportionation, three-body association, and
thermal decomposition. The full set of reactions consists, in
fact, of 941 reversible reactions written in the forward and
backward directions. For most of them, the rate constant of the
backward reaction is calculated applying the principle of detailed
balance and using the rate constant of the forward process and
thermochemical data of the species involved. This chemical network
ensures that the chemical composition will evolve towards a
chemical equilibrium state (see Venot et al. 2012). The set of
rate constants and thermochemical data of our mechanism comes from
combustion chemistry. More specifically, it consists of a C/H/O
reaction base developed for industrial applications (Fournet et
al. 1999; Bounaceur et al. 2010) which has been validated for
species containing up to 2 carbon atoms and a nitrogen base
constructed to deal with NO$_x$ compounds and other
nitrogen-containing species (Coppens et al. 2007; Konnov 2009).
The mechanism has been validated against numerous combustion
experiments over a wide range of temperature (300--2500 K) and
pressure (0.01--100 bar), and it has been found suitable to model
the atmospheric chemistry of hot Jupiters (Venot et al. 2012).

There are two main differences with respect to the original
nominal mechanism used in Venot et al. (2012). First, in this
study we also model the nightside of HD 209458b whose upper
atmospheric layers may have temperatures lower than 300 K, i.e.
outside the temperature range of validation of the chemical
network. At these low temperatures the rate constants of a few
reactions, mostly backward reactions whose rate constant is
calculated applying detailed balance, reach too large values.
For these few reactions we have, either adopted a
different rate constant expression from the literature, or imposed
a maximum rate constant (2 $\times$ 10$^{-9}$ cm$^3$ s$^{-1}$ and
10$^{-28}$ cm$^6$ s$^{-1}$ for bimolecular and termolecular
reactions, respectively) to get reasonable values down to
temperatures lower than 300 K. The impact of these modifications
on the abundances of the major species is, nonetheless,
negligible. A second difference is that in this study we have not
considered photochemical processes, so that the excited states of
oxygen and nitrogen atoms, O($^1$D) and N($^2$D), have not been
included, as they are mainly produced by photodissociation of
various molecules. Previous studies have found that photochemistry
has a very minor impact on the atmospheric composition of the
dayside of HD 209458b, due to the high temperatures which help to
maintain chemical equilibrium (Moses et al. 2011; Venot et al.
2012). The vertical temperature profiles adopted by these authors
above the 100 bar layer are mainly based on the results of the
three-dimensional circulation model of Showman et al. (2009),
which results in dayside temperatures higher than those found with
our time-dependent radiative model. We therefore note that the
lower temperatures we find are likely to increase the impact of
photochemistry, as compared with the studies of Moses et al.
(2011) and Venot et al. (2012), on the abundances of some species,
noticeably HCN, CO$_2$, CH$_4$, and NH$_3$, in the upper (above
the 10$^{-4}$--10$^{-5}$ bar layer) dayside atmosphere of HD
209458b. At these altitudes horizontal winds are, anyway,
substellar-to-antistellar rather than superrotating.

Some of the limitations of our current chemical modelling
approach, where layers are assumed to be independent from each
other (they are neither radiatively coupled through the transfer
of ultraviolet photons nor dynamically coupled by vertical
mixing), will be addressed in a future study.

\section{Results and discussion} \label{sec-results}

\subsection{Zonal wind: effects on the chemistry} \label{subsec-chemistry-superrotation}

The impact of a uniform zonal wind, mimicking an equatorial
superrotation, on the atmospheric chemistry of a hot Jupiter such
as HD 209458b may be understood in terms of timescales. On the one
hand, we have the chemical timescale of the different chemical
transformations that take place as the wind moves material between
the hot dayside and the cooler nightside, and on the other hand,
there is the dynamical timescale related to the zonal wind motion.

\begin{figure}
\centering
\includegraphics[angle=0,width=0.92\columnwidth]{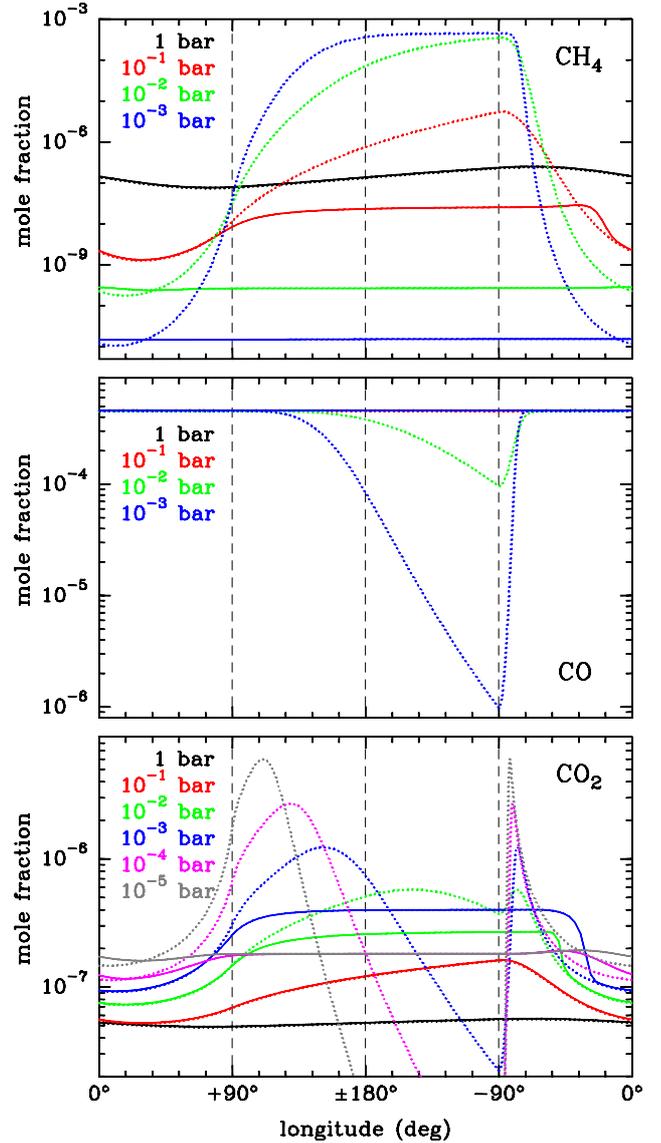}
\caption{Longitudinal distributions of the mole fractions of
CH$_4$, CO, and CO$_2$ for selected pressure levels, as calculated
by the chemical kinetics model with a uniform zonal wind of 1 km
s$^{-1}$ (solid lines) and by chemical equilibrium adopting the
longitudinal temperature profile of each pressure level (dotted
lines).} \label{fig-abundances-horizontal}
\end{figure}

If the atmosphere were at chemical equilibrium everywhere, we
would expect important differences in the chemical composition of
the dayside and nightside due to the high temperature contrast
between both planet sides. For example, at chemical equilibrium
most of the carbon is in the form of CO at high temperatures and
as CH$_4$ at temperatures below $\sim$1000 K. This is illustrated
in Fig~\ref{fig-abundances-horizontal}, where chemical equilibrium
abundances of a few molecules (CH$_4$, CO, and CO$_2$), as
computed along the longitudinal temperature profile of each
pressure level, are shown (as dotted lines) as a function of
longitude for selected pressures. It is seen that for $p$ $\leq$
10$^{-2}$ bar, methane has a low abundance in the dayside, where
most of carbon is in the form of CO, but becomes very abundant in
the nightside at the expense of carbon monoxide.

The presence of a zonal wind, however, produces significant
departures from chemical equilibrium. In
Fig~\ref{fig-abundances-horizontal} we also show (as solid lines)
the mole fractions of CH$_4$, CO, and CO$_2$ resulting from the
chemical kinetics model after 10 rotation cycles, when the
abundances of the major species have already reached a periodic
state. We can identify two major types of regimes. In the hot and
dense bottom layers of the atmosphere chemical timescales are
short and chemical equilibrium is attained (see e.g. the case of
CH$_4$ at 1 bar). At higher altitudes, however, temperatures and
pressures decrease and chemical timescales become longer than the
dynamical timescale, producing a "quenching" effect on the
abundances (see e.g. the case of CH$_4$ at 10$^{-2}$ and 10$^{-3}$
bar). In these upper layers, species tend to reach an abundance
uniform with longitude, close to the chemical equilibrium values
of the hottest dayside regions. This latter fact makes very
convenient to start the integration with the chemical equilibrium
composition at the substellar meridian, in which case the initial
abundances are already close to the final quenched values and so
the convergence is considerably accelerated. We have verified the
non-dependence of our results on the initial chemical composition
by starting with chemical equilibrium compositions at other
longitudes, such as the antistellar meridian, and we find that the
same periodic state is reached for the abundances of the major
species, although at much longer integration times, especially in
low pressure layers.

\begin{figure*}
\centering
\includegraphics[angle=-90,width=0.98\textwidth]{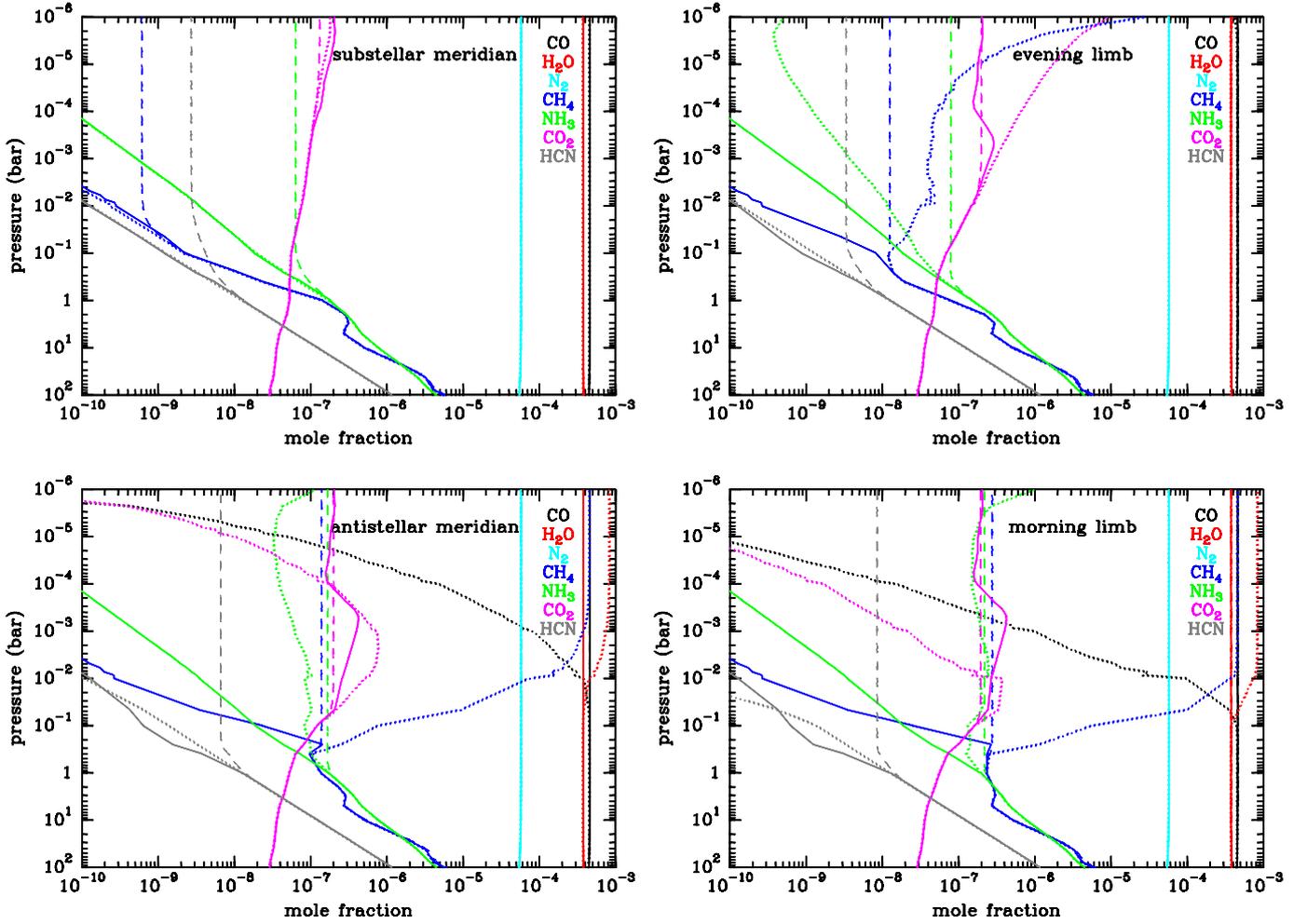}
\caption{Vertical distributions of the major species, after H$_2$
and He, at 4 longitudes: substellar meridian, evening limb,
antistellar meridian, and morning limb. We show the mole fractions
given by the zonal wind thermochemical kinetics model (solid
lines), by a one dimensional model in the vertical direction
including thermochemical kinetics and vertical eddy diffusion
(dashed lines), and by chemical equilibrium adopting the ($p,T$)
vertical profile at each longitude (dotted lines). Note
that abundances above the 10$^{-4}$--10$^{-5}$ bar level may not
be reliable due to the lack of photochemistry in our model and, in
the case of the zonal wind model, to a likely change in the
circulation pattern.} \label{fig-abundances-vertical}
\end{figure*}

In the layers of transition between these two regimes the
behaviour of the chemical composition with longitude is such that
abundances are close to the chemical equilibrium values in the hot
dayside, where chemical kinetics proceeds faster, and remain
somewhat quenched in the cooler nightside, where chemical kinetics
is slower (see e.g. the case of CH$_4$ at 10$^{-1}$ bar). Each
species has its own chemical timescale of formation and
destruction, and so the transition between the chemical
equilibrium and quenching regimes is located at a different
pressure level for each species. For CH$_4$ and CO the quenching
region is found above the 10$^{-1}$ bar level (see
Fig~\ref{fig-abundances-horizontal}), so that the large
day-to-night abundance variations predicted by chemical
equilibrium for these two molecules at $p$ $\leq$ 10$^{-2}$ bar
are not effectively attained. This result was also found by Cooper
\& Showman (2006) using a simple conversion scheme for CO and
CH$_4$. For other molecules with shorter chemical timescales the
quenching level is located at higher altitudes (above the
10$^{-3}$--10$^{-4}$ bar level for CO$_2$; see
Fig~\ref{fig-abundances-horizontal}).

We have adopted an equatorial wind velocity of 1 km s$^{-1}$,
although three dimensional circulation models of HD 209458b find
equatorial wind velocities in the range 0.1--10 km s$^{-1}$
(Showman et al. 2009; Heng et al. 2011; Miller-Ricci Kempton \&
Rauscher 2012). The wind velocity affects the distribution of the
abundances in two ways. On the one hand, it has an impact on the
longitudinal temperature profile; the slower the wind velocity the
higher the day-to-night temperature contrast (see Iro et al.
2005). It therefore affects the chemical equilibrium composition
attained in the bottom atmospheric layers, which depends on the
local temperature. On the other hand, the wind velocity determines
the horizontal dynamical timescale and therefore the quenching
level, which is shifted upwards with slower wind velocities.
Although our results depend to some extent on the adopted wind
velocity, overall the qualitative behaviour of the distribution of
the abundances remains the same.

The zonal wind, therefore, tends to homogenize the chemical
composition of the atmosphere, bringing species abundances in the
nightside close to those prevailing in the dayside. In the upper
atmospheric layers, above the 0.1--10$^{-4}$ bar level depending
on the species, abundances are quenched to chemical equilibrium
values of the hottest dayside longitudes. Below this quenching
level, where chemical timescales become comparable or shorter than
horizontal dynamical timescales, abundances can experience
significant variations with longitude induced by the longitudinal
temperature gradients. At these pressure levels the abundances of
some molecules can be quite different in the two meridians of the
planet's terminator (the morning and evening limbs), with
consequences for the interpretation of transmission spectra (see
below). Deeper in the atmosphere, below the 1 bar level,
temperatures are high and uniform with longitude, so that the
chemical composition is also uniform with longitude and given by
chemical equilibrium.

At high enough levels, horizontal quenching makes the chemical
composition to be homogeneous with longitude, which allows to
apply retrieval methods based on a single vertical chemical
profile when interpreting transmission spectra. At these
chemically homogenous levels, the molecular composition of the
limbs is nearly independent on the local temperature and is mainly
determined by the substellar temperature. This raises an
interesting question: can the abundances measured at the limbs be
used as a thermometer for the substellar region$?$ Although they
can indeed provide constraints on the hottest temperatures
experienced by the gas, quenching occurs at a different
temperature for each species. Therefore, not only the observed
mixture is in disequilibrium with the local temperature but it is
also in disequilibrium with any temperature. For instance, at the
10$^{-4}$ bar level, the sunrise and sunset limbs are respectively
at 510 and 820 K and they both exhibit the same abundances for CO,
CH$_4$, CO$_2$, N$_2$ and NH$_3$. The CH$_4$/CO and NH$_3$/N$_2$
abundance ratios both correspond to a chemical equilibrium
composition at 1300~K, which is nearly the temperature of the
substellar region, but the CO$_2$/CO abundance ratio corresponds
to an equilibrium composition of a cooler mixture (at 1145 K). Of
course these quenching temperatures depend on the elemental
composition and the wind velocity, and are derived from a rather
simple circulation pattern. In addition, we should keep in mind
that composition may vary significantly between the equator and
the poles. We will come back to this point at the end of
Sec.~\ref{subsec-chemistry-horizontal-vertical}.

\subsection{Horizontal vs vertical quenching} \label{subsec-chemistry-horizontal-vertical}

In the limit of a solid body rotating atmosphere, where dynamics
is restricted to a uniform zonal wind, molecular abundances are
quenched above a certain pressure level to chemical equilibrium
dayside values. The distribution of the chemical composition
computed in this way is, however, likely to be unrealistic due to
the existence of vertical transport processes, which tend to
quench abundances vertically above a certain height to chemical
equilibrium values characteristic of hot deep regions (Moses et
al. 2011; Venot et al. 2012).

In order to compare the predictions of our zonal wind model with a
model where dynamics is restricted to vertical mixing, we have
constructed a one dimensional model in the vertical direction that
includes thermochemical kinetics and vertical eddy diffusion, and
applied it to various longitudes. Photochemistry is not included
to allow for a proper comparison with the zonal wind model. We use
a code which is based on the same principles of that used by
Dobrijevic et al. (2010) and Venot et al. (2012). We adopt the
temperature-pressure vertical profiles calculated at each
longitude with the time-dependent radiative model (see bottom
panel of Fig~\ref{fig-tk-distribution}) and the chemical network
described in Sec.~\ref{subsec-model-chemistry}. The
dayside-averaged eddy diffusion coefficient profile estimated by
Moses et al. (2011) from the three dimensional circulation model
of Showman et al. (2009) is adopted, for simplicity, at every
longitude.

In Fig~\ref{fig-abundances-vertical} we show the vertical
distribution of the major species at the substellar meridian,
evening limb, antistellar meridian, and morning limb, as computed
by the zonal wind model (solid lines), by the vertical mixing
model (dashed lines), and by chemical equilibrium (dotted lines).
Both the zonal wind and vertical mixing models predict important
departures from chemical equilibrium, especially in the cool
nightside regions (see e.g. CO and CH$_4$ at the antistellar
meridian). The vertical distribution of the abundances predicted
by the zonal wind model is quite similar at the 4 longitudes, and
is essentially given by the chemical equilibrium composition
calculated at the substellar meridian. There are some differences
in the vertical abundance profile of some molecules when moving
from one longitude to another (see e.g. CO$_2$ and CH$_4$ at the
substellar and antistellar meridians) although the main effect of
the zonal wind is that it tends to homogenize the chemical
composition at the limbs and nightside to chemical equilibrium
dayside values. The vertical mixing model shows a quite different
picture, where abundances remain at chemical equilibrium in the
deep atmosphere, and are quenched above the 1--0.01 bar pressure
level, depending on the molecule, to the chemical equilibrium
values characteristic of this quenching level. Since the vertical
temperature profile is different at each longitude, the value at
which the abundance of a given molecule is quenched can also vary
from one longitude to another (see e.g. CH$_4$ at the
substellar and antistellar meridians). We note that photochemistry
may play an important role above the 10$^{-4}$--10$^{-5}$ bar
level, so that the abundances calculated by the zonal wind and
vertical mixing models above this layer may not be very reliable.

Therefore, in the limit of a zonal wind, abundances are quenched
horizontally (above the 0.1--10$^{-4}$ bar level) to chemical
equilibrium dayside values, while in the limit where dynamics is
restricted to vertical mixing, abundances are quenched vertically
(above the 1--0.01 bar level) to chemical equilibrium values
characteristic of the hot deep layers. We can get a rough idea of
the relative importance of horizontal and vertical quenching from
timescales arguments. The dynamical timescale of horizontal mixing
$\tau_{{\rm dyn},h}$ may be roughly estimated as $\pi R_p/u$,
where $R_p$ is the radius of the planet, $u$ is the zonal wind
speed, and where the distance between substellar and antistellar
points has been adopted as length scale. For our zonal wind speed
of 1 km s$^{-1}$ we get $\tau_{{\rm dyn},h}$ $\sim$ 3 $\times$
10$^{5}$ s. Similarly, a rough estimate of the dynamical timescale
of vertical mixing $\tau_{{\rm dyn},v}$ may be obtained as
$H^2/K_{zz}$, where $H$ is the atmospheric scale height and
$K_{zz}$ is the eddy diffusion coefficient. Adopting values for
$H$ and $k_{zz}$ in the 1 bar to 1 mbar pressure range of HD
209458b's atmosphere ($H$ = 300--700 km,  $K_{zz}$ =
10$^{10}$--10$^{11}$ cm$^2$ s$^{-1}$) we find that $\tau_{{\rm
dyn},v}$ takes values between 10$^4$ and 5 $\times$ 10$^5$ s. We
should keep in mind that the above timescale values are just a
crude estimation. On the one hand, the zonal wind speed may span a
broad range of values up to 10 km s$^{-1}$ depending on height and
latitude, according to three dimensional circulation models of HD
209458b (Showman et al. 2009; Heng et al. 2011; Miller-Ricci
Kempton \& Rauscher 2012). On the other hand, the use of the
atmospheric scale height to estimate the dynamical timescale for
vertical mixing may not be an accurate choice, as discussed by
Smith (1998). Finally, the values of the eddy diffusion
coefficient adopted, which are estimated by Moses et al. (2001)
from the three dimensional circulation model of Showman et al.
(2009) as the horizontally averaged global rms vertical velocity
times the height scale, may be quite different if they are more
properly estimated through, e.g., the diffusive behaviour of a
passive tracer included in the circulation model. Moreover,
dynamical timescales are expected to experience important
variations as a function of longitude, latitude, and height.
Taking into account all these limitations, we may just conclude
that the dynamical timescales for horizontal and vertical mixing
have the same order of magnitude. It is, therefore, very likely
that both horizontal and vertical quenching are simultaneously at
work in the atmosphere of HD 209458b, which may result in a
complex atmospheric distribution of the abundances of some
species. Our timescales estimates are similar to those
found by Cooper \& Showman (2006) for HD 209458b, although these
authors conclude that horizontal quenching is not important.
Cooper \& Showman (2006) argue that chemical timescales exceed the
horizontal dynamical timescale above the 3 bar level, where
abundances have already been quenched by vertical mixing. It is,
nevertheless, the comparison between the dynamical timescales for
horizontal and vertical mixing, $\tau_{{\rm dyn},h}$ and
$\tau_{{\rm dyn},v}$, which allows to argue on the dominant
(either horizontal or vertical) quenching mechanism, whenever
$\tau_{\rm chem}$ remains longer than any dynamical timescale.

\begin{figure}
\centering
\includegraphics[angle=-90,width=0.99\columnwidth]{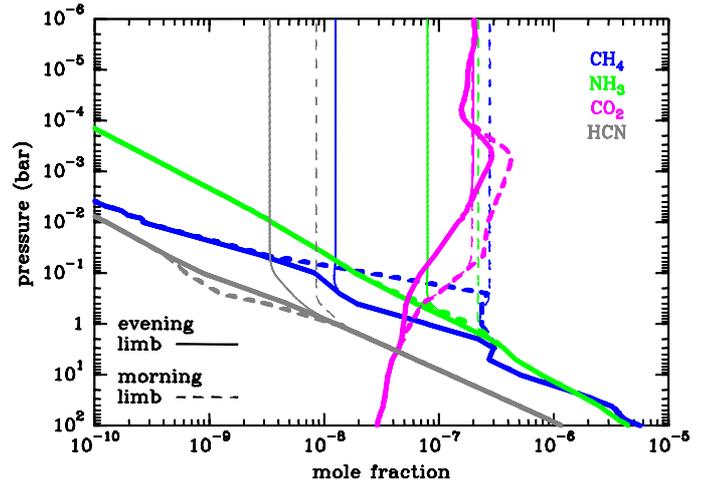}
\caption{Vertical distributions of the abundances of CH$_4$,
NH$_3$, CO$_2$, and HCN at the evening (solid lines) and morning
(dashed lines) limbs, as calculated by the zonal wind model (thick
lines) and by the vertical mixing model (thin lines). Note
that abundances above the 10$^{-4}$--10$^{-5}$ bar level may not
be reliable due to the lack of photochemistry in our model and, in
the case of the zonal wind model, to a likely change in the
circulation pattern.} \label{fig-abundances-limbs}
\end{figure}

As shown in Fig~\ref{fig-abundances-vertical}, the molecules CO,
H$_2$O, and N$_2$ (as well as H$_2$ which is not shown) show a
uniform abundance with height and longitude, either adopting the
zonal wind model or the vertical mixing model. For these molecules
it is, therefore, of no relevance whether horizontal or vertical
quenching dominates. The vertical abundance profile of the other
major molecules CH$_4$, NH$_3$, CO$_2$, and HCN shows important
differences when calculated with the zonal wind model or with the
vertical mixing one. Moreover, using either of the two models,
some abundance variations with longitude are found. These
molecules, therefore, may have different abundances in the evening
and morning limbs, which would have consequences for the
interpretation of transmission spectra. In
Fig~\ref{fig-abundances-limbs} we compare the vertical
distribution of the abundances of CH$_4$, NH$_3$, CO$_2$, and HCN
at the evening and morning limbs, as computed with the zonal wind
model and the vertical mixing model. In the case of CH$_4$ we see
that abundance differences up to one order of magnitude are found
between the two limbs, either in the 1--0.1 bar pressure regime
(if horizontal mixing dominates) or at every height above the 1
bar level (if vertical mixing dominates). NH$_3$ and HCN show a
similar behaviour, with very small abundance differences in the
1--0.1 bar pressure range, as predicted by the zonal wind model,
and a difference of a factor of 2--3 above the 1 bar level,
according to the vertical mixing model. Finally, in the case of
CO$_2$ it is predicted that a moderate abundance difference
between the two limbs would be found if horizontal mixing
dominates, while no difference would be observed if it is vertical
mixing which dominates. Methane seems, therefore, to be the
species that could show the most important abundance differences
between the evening and morning limbs.

The retrieval of atmospheric properties and composition from
primary transits implies the comparison of the observed
transmission spectrum with a synthetic one obtained with an
atmosphere model. Ideally, this model would be a detailed three
dimensional model coupling dynamics, radiative transfer, and
chemistry, so that the only parameters to be adjusted would be the
elemental abundances. This is unfortunately not doable at this
time and one dimensional models are currently used for this
purpose. Our study shows that the one dimensional approach is
acceptable because of the longitudinal homogenization produced by
zonal circulation. On the other hand, transmission spectra
generally probes simultaneously both west and east
limbs\footnote{Ingress and egress could in principle be treated
separately, although in the practice this is very difficult due
to, among other reasons, stellar limb darkening effects.}, which
have very different temperatures. Since transmission spectroscopy
is very sensitive to the temperature via the scale height (Tinetti
et al. 2007), retrieval should include the contribution of (at
least) two different vertical temperature profiles. Transmission
spectra probes also latitudes from the equator to the poles, and
therefore latitudinal temperature variations are an additional
complication.

Another important matter of debate is the lack of self-consistency
between chemical and temperature vertical profiles used by one
dimensional retrieval models. It would be tempting to use one
dimensional profiles in which molecular abundances and temperature
are computed consistently, in order to limit the dimensions of the
parameter space that has to be explored when the model is not
self-consistent and any composition is allowed (e.g. Tinetti et
al. 2007; Swain et al. 2008; Beaulieu et al. 2011). We have shown,
however, that the temperature variations experienced by the gas in
the zonal flow affect the composition so that abundances above the
1 bar layer are determined by a mixture of temperatures, which are
close to those of the hottest regions encountered rather than to
the local temperature. Because this effect competes with the
chemical quenching produced by vertical mixing, a minimum
self-consistent approach would imply to calculate the limb
vertical structure and composition with a time dependent model
that accounts for both the vertical mixing and the horizontal
circulation (an additional complication would be to account for
latitudinal chemical gradients). This will be the next upgrade of
our model. To our knowledge, such model has never been used to
interpret spectra, although it may help to shed light on some
unexplained observations and contradictory analyses.

\section{Summary}

In this study we have carried out an attempt to understand the
impact of atmospheric circulation on the distribution of the
atmospheric constituents of a tidally locked hot Jupiter, such as
HD 209458b. The temperature structure and the distribution of the
chemical composition as a function of longitude and height have
been computed in the limit of a solid body rotating atmosphere,
which mimics a uniform zonal wind. Adopting an equatorial wind
speed of 1 km s$^{-1}$, we find that chemical equilibrium is
attained below the 0.1--10$^{-4}$ bar pressure level, depending on
the molecule, while above this transition layer, molecular
abundances are quenched to chemical equilibrium values
characteristic of the hottest dayside regions. Reasoning based on
timescales arguments indicate that such horizontal quenching is
likely to compete in HD 209458b's atmosphere with the vertical
quenching induced by eddy diffusion processes, implying that in
some atmospheric regions molecular abundances are quenched
horizontally to dayside values while in other regions abundances
are quenched vertically to chemical equilibrium values
characteristic of deep layers. Moderate abundance variations
between the evening and morning limbs are found for some
molecules, up to one order of magnitude in the case of CH$_4$,
assuming that dynamics is either restricted to a zonal wind or to
vertical mixing, which may have consequences for the
interpretation of transmission spectra obtained during primary
transits.

Ideally, a realistic distribution of the molecular abundances
throughout the whole planetary atmosphere could be obtained by
coupling a three dimensional circulation model to a robust
chemical network. Such approach is, however, a complicated task
for numerical reasons. For the moment, insights into the influence
of atmospheric dynamics on the chemistry must be addressed either
coupling a three dimensional circulation model to a simple
chemical kinetics scheme, as done by Cooper \& Showman (2006), or
by using a simplified dynamical model and a robust chemical
network, as done in this study. We plan to overcome some of the
limitations of our current modelling approach, which does not
consider neither photochemistry nor the coupling between
horizontal and vertical mixing, in a future study.

\begin{acknowledgements}

M.A., O. V., F. S., and E. H. acknowledge support from the
European Research Council (ERC Grant 209622: E$_3$ARTHs). Computer
time for this study was provided by the computing facilities MCIA
(M\'esocentre de Calcul Intensif Aquitain) of the Universit\'e de
Bordeaux and of the Universit\'e de Pau et des Pays de l'Adour. We
thank the anonymous referee for a constructive report that helped
to improve this manuscript.

\end{acknowledgements}


\begin{thebibliography}{}

\bibitem[Asplund et al.]{asp2009} Asplund, M., Grevesse, N., Sauval, A. J., \& Scott, P. 2009, \araa, 47, 481
\bibitem[Batalha et al.]{bat2012} Batalha, N. M., Rowe, J. F., Bryson, S. T., et al. 2012, \texttt{arXiv1202.5852}
\bibitem[Beaulieu et al.]{bea2011} Beaulieu, J.-P., Tinetti, G., Kipping, D. M., et al. 2011, \apj, 731, 16
\bibitem[Bounaceur et al.]{bou2010} Bounaceur, R., Herbinet, O., Fournet, R., et al. 2010, SAE Technical Paper 2010-01-0546
\bibitem[Charbonneau et al.]{cha2000} Charbonneau, D., Brown, T. M., Latham, D. W., \& Mayor, M. 2000, \apj, 529, L45
\bibitem[Charbonneau et al.]{cha2002} Charbonneau, D., Brown, T. M., Noyes, R. W., \& Gilliland, R. L. 2002, \apj, 568, 377
\bibitem[Cho et al.]{cho2003} Cho, J. Y.-K., Menou, K., Hansen, B. M. S., \& Seager, S. 2003, \apj, 587, 117
\bibitem[Cho et al.]{cho2008} Cho, J. Y.-K., Menou, K., Hansen, B. M. S., \& Seager, S. 2008, \apj, 675, 817
\bibitem[Cooper \& Showman]{coo2005} Cooper, C. S. \& Showman, A. P. 2005, \apj, 629, L45
\bibitem[Cooper \& Showman]{coo2006} Cooper, C. S. \& Showman, A. P. 2006, \apj, 649, 1048
\bibitem[Coppens et al.]{cop2007} Coppens, F. H. V., De Ruyck, J., Konnov, A. A. 2007, \cflame, 149, 409
\bibitem[Dobbs-Dixon et al.]{dob2012} Dobbs-Dixon, I., Agol, E., \& Burrows, A. 2012, \apj, 751, 87
\bibitem[Dobrijevic et al.]{dob2010} Dobrijevic, M., Cavali\'e, T., H\'ebrard, E., et al. 2010, \pss, 58, 1555
\bibitem[Gordon \& McBride]{gor1994} Gordon, S., \& McBride, B. J. 1994, NASA Reference Publication 1311, I
\bibitem[Fournet et al.]{fou1999} Fournet, R., Baug\'e, J. C., \& Battin-Leclerc, F. 1999, \ijck, 31, 361
\bibitem[Grillmair et al.]{gri2008} Grillmair, C. J., Burrows, A., Charbonneau, D., et al. 2008, \nature, 456, 767
\bibitem[Guillot et al.]{gui1996} Guillot, T., Burrows, A., Hubbard, W. B., et al. 1996, \apj, 459, L35
\bibitem[Guillot \& Showman]{gui2002} Guillot, T. \& Showman, A. P. 2002, \aap, 385, 156
\bibitem[Heng et al.]{hen2011} Heng, K., Menou, K., \& Phillips, P. J. 2011, \mnras, 413, 2380
\bibitem[Henry et al.]{hen2000} Henry, G. W., Marcy, G. W., Butler, R. P. \& Vogt, S. S. 2000, \apj, 529, L41
\bibitem[Iro et al.]{iro2005} Iro, N., B\'ezard, B., \& Guillot, T. 2005, \aap, 436, 719
\bibitem[Knutson et al.]{knu2007} Knutson, H. A., Charbonneau, D., Allen, L. E., et al. 2007, \nature, 447, 183
\bibitem[Konnov]{kon2009} Konnov, A. A. 2009, \cflame, 156, 2093
\bibitem[Kopparapu et al.]{kop2012} Kopparapu, R. K., Kasting, J. F., \& Zahnle, K. J. 2012, \apj, 745, 77
\bibitem[Line et al.]{lin2010} Line, M. R., Liang, M. C., \& Yung, Y. L. 2010, \apj, 717, 496
\bibitem[Madhusudhan et al.]{mad2011} Madhusudhan, N., Harrington, J., Stevenson, K. B., et al. 2011, \nature, 469, 64
\bibitem[Mayor \& Queloz]{may1995} Mayor, M. \& Queloz, D. 1995, \nature, 378, 355
\bibitem[Mazeh et al.]{maz2000} Mazeh, T., Naef, D., Torres, G., et al. 2000, \apj, 532, L55
\bibitem[Menou \& Rauscher]{men2009} Menou, K. \& Rascher, E. 2009, \apj, 700, 887
\bibitem[Miller-Ricci Kempton \& Rauscher]{mil2012} Miller-Ricci Kempton, E. \& Rauscher, E. 2012, \apj, 751, 117
\bibitem[Moses et al.]{mos2011} Moses, J. I., Visscher, C., Fortney, J. J., et al. 2011, \apj, 737, 15
\bibitem[Rothman et al.]{rot2009} Rothman, L. S., Gordon, I. E., Barbe, A., et al. 2009, J. Quant. Spec. Radiat. Transf., 110, 533
\bibitem[Rothman et al.]{rot2010} Rothman, L. S., Gordon, I. E., Barber, R. J., et al. 2010, J. Quant. Spec. Radiat. Transf., 111, 2139
\bibitem[Showman \& Guillot]{sho2002} Showman, A. P. \& Guillot, T. 2002, \aap, 385, 166
\bibitem[Showman et al.]{sho2008} Showman, A. P., Cooper, C. S., Fortney, J. J. \& Marley, M. S. 2008, \apj, 682, 559
\bibitem[Showman et al.]{sho2009} Showman, A. P., Fortney, J. J., Lian, Y., et al. 2009, \apj, 699, 564
\bibitem[Showman et al.]{sho2012} Showman, A. P., Fortney, J. J.,Lewis, N. K., \& Shabram, M. 2012, \texttt{arXiv1207.5639}
\bibitem[Sing et al.]{sin2009} Sing, D. K., D\'esert, J.-M., Lecavelier des Etangs, A., et al. 2009, \aap, 505, 891
\bibitem[Smith]{smi1998} Smith, M. D. 1998, \icarus, 132, 176
\bibitem[Snellen et al.]{sne2010} Snellen, I. A. G., de Kok, R. J., de Mooij, E. J. W., \& Albrecht, S. 2010, \nature, 465, 1049
\bibitem[Southworth]{sou2010} Southworth, J. 2010, \mnras, 408, 1689
\bibitem[Swain et al.]{swa2008} Swain, M. R., Vasisht, G., Tinetti, G., et al. 2008, \nature, 452, 329
\bibitem[Swain et al.]{swa2009} Swain, M. R., Tinetti, G., Vasisht, G., et al. 2009, \apj, 704, 1616
\bibitem[Tinetti et al.]{tin2007} Tinetti, G., Vidal-Madjar, A., Liang, M.-C., et al. 2007, \nature, 448, 169
\bibitem[Venot et al.]{ven2012} Venot, O., H\'ebrard, E., Ag\'undez, M., et al. 2012, \aap, 546, 43
\bibitem[Wakelam et al.]{wak2012} Wakelam, V., Herbst, E., Loison, J.-C., et al. 2012, \apjs, 199, 21
\bibitem[Zahnle et al.]{zah2009} Zahnle, K., Marley, M. S., Freedman, R. S., et al. 2009, \apj, 701, L20

\end{thebibliography}
\end{document}